\documentstyle[12pt]{article}
\huge
\renewcommand{\theequation}{\arabic{section}.\arabic{equation}} 
\def\setzero{\setcounter{equation}{0}}

\newcounter{eqalph}
\def\bph{\setcounter{eqalph}{\value{equation}}
   \addtocounter{eqalph}{1}
   \setcounter{equation}{0}
   \renewcommand{\theequation}{\arabic{section}.\arabic{eqalph}\alph{equation}}}
\def\eph{\setcounter{equation}{\value{eqalph}}
   \renewcommand{\theequation}{\arabic{section}.\arabic{equation}}
\par\noindent}

\begin{document}

\baselineskip 18pt

\def \sech{{\rm sech}}
\def \tanh{{\rm tanh}}
\def \cn{{\rm cn}}
\def \sn{{\rm sn}}
\def\bm#1{\mbox{\boldmath $#1$}}
\newfont{\bg}{cmr10 scaled\magstep4}
\newcommand{\bzr}{\smash{\hbox{\bg 0}}}
\newcommand{\bzl}{%
   \smash{\lower1.7ex\hbox{\bg 0}}}
\title{Five-Dimensional Gauge Theories  \\and 
\\Whitham-Toda Equation} 
\date{\today}
\author{ Masato {\sc Hisakado}  
\\
\bigskip
\\
{\small\it Department of Pure and Applied Sciences,}\\
{\small\it University of Tokyo,}\\
{\small\it 3-8-1 Komaba, Meguro-ku, Tokyo, 153, Japan}}
\maketitle

\vspace{20 mm}

Abstract


The  five-dimensional supersymmetric $SU(N)$
 gauge theory is studied in the framework  
of the relativistic Toda chain.
This equation  can be embedded in two-dimensional Toda lattice hierarchy.
This system  has the conjugate  
structure.
This conjugate structure corresponds to  the charge conjugation.

\vfill
\par\noindent
{\bf  }

\newpage

\section{Introduction}
Recently there has been a tremendous  progress in understanding the 
non-perturvative behavior of supersymmetric gauge theories in 
four dimensions.\cite{sw}
The exact low energy effective action  was determined  for a large class
of $N=1$ and $N=2$ theories.
The main tool  for such determination is the use of holomorphy 
 of various functions and the electric-magnetic duality.
They constraint the moduli space.

The solution of the pure $N=2$ super Yang Mills theory 
originally is obtained by Seiberg and Witten  and 
and reinterpreted by Grosky et al. in terms of integrable systems.
The correspondence between the Seiberg-Witten solution 
and  the elliptic 
Whitham-KdV equation is pointed out.\cite{gk}
The observation was soon extended to the solution for the 
other classical gauge groups.
Martinec and Warner noticed that these hyper elliptic curves
 are spectral curves of  affine Toda chain systems.\cite{mw}
The prepotential ${\cal F}$ is  a Whitham type $\tau$ function.\cite{nt}

Consider a five-dimensional supersymmetric gauge theory.
Take the space-time manifold to be $M=X\times S_{1}$,
 where $X$ is a four manifold and $S_{1}$ is a circle.
Nekrasov pointed out  that the solution 
of five-dimensional theory can be obtained  by the 
relativistic Toda chain (RTC) and determined  the exact non-perturvative 
prepotential of the five-dimensional theory.\cite{n}
Recently the relation between five-dimensional gauge theory with matters
 and Calabi-Yau geometry are studied.\cite{i}

Karchev et  al. demonstrated that 
RTC is a special reduction of the two-dimensional Toda 
Lattice hierarchy.\cite{m}
We  extend this system to the periodic case.\cite{r},\cite{br}
Since the system has  two Lax operators, 
we can obtain  two curves.
However the $\tau$-function  is one.
To describe the five-dimensional moduli space
 the two curves have to be unified.
 
This letter is organized as follows.
In  section 2 we discuss  periodic RTC and 
obtain the two curves.
This system has the conjugate structure.
In section 3 we study it from the viewpoint  of the
 Whitham-Toda hierarchy.\cite{w}-\cite{bk}
 We can see the inhomogeneity 
of the two times $t_{k}$ and $\bar{t}_{k}$.
In section 4 we apply it to  the five-dimensional gauge theory.
The last section is devoted to the concluding remarks.

\setzero

\section{Lax representation for RTC}
We consider the  eigenfunctions $\Phi_{n}$ and $\Phi^{*}_{n}$.
These eigen functions satisfy the following recurrent 
relations which give  the Lax operators: \cite{m}
\begin{eqnarray}
\Phi_{n+1}-\frac{S_{n}}{S_{n-1}}\Phi_{n}
&=&
z[\Phi_{n}-\frac{S_{n}}{S_{n-1}}(1-S_{n}S^{*}_{n-1})\Phi_{n-1}],
\nonumber \\
\Phi^{*}_{n+1}-\frac{S_{n}^{*}}{S_{n-1}^{*}}\Phi^{*}_{n}
&=&
z^{-1}[\Phi^{*}_{n}-\frac{S^{*}_{n}}{S^{*}_{n-1}}(1-S_{n}S^{*}_{n-1})
\Phi^{*}_{n-1}].
\end{eqnarray}
with
\begin{equation}
S_{n}=S_{n+N}, \;\;\;\; S_{n}^{*}=S_{n+N}^{*}.
\label{pc}
\end{equation}
The superscript $*$ does not mean the complex conjugation.
We define the next  function:
\begin{equation}
\frac{h_{n+1}}{h_{n}}=1-S_{n}S_{n}^{*}.
\label{hs}
\end{equation}
In terms of the canonically conjugate variables for 
the RTC
$(q_{n}, p_{n}, p_{n}^{*})$,
 we can immediately read off
\begin{equation}
\frac{S_{n}}{S_{n-1}}=-\exp(\epsilon p_{n}),\;\;\;
\frac{S_{n}^{*}}{S_{n-1}^{*}}=-\exp(\epsilon p_{n}^{*}),
\end{equation}
\begin{equation}
\frac{h_{n}}{h_{n-1}}=\epsilon^{2}\exp(q_{n}-q_{n-1}),
\end{equation}
with the periodic condition 
\begin{equation}
q_{n+N}=q_{n},\;\;\;p_{n+N}=p_{n},\;\;\;
p_{n+N}^{*}=p_{n}^{*}.
\end{equation}
Adding this  the canonical coordinates satisfy the 
constraint
\begin{equation}
\sum_{i}q_{i}=\sum_{i}p_{i}=\sum_{i}p_{i}^{*}=0.
\label{ic}
\end{equation}

We next consider the equations which describe the time dependence
of $\Phi_{n}$ and $\Phi_{n}^{*}$.
 For example $t_{1}$ and $\bar{t}_{1}$ give  the following evolution equations:
\bph
\begin{equation}
\frac{\partial \Phi_{n}(z)}{\partial t_{1}}
=-\frac{S_{n}}{S_{n-1}}\frac{h_{n}}{h_{n-1}}
(\Phi_{n}(z)-z\Phi_{n-1}),
\end{equation}
\begin{equation}
\frac{\partial \Phi_{n}(z)}{\partial \bar{t}_{1}}
=\frac{h_{n}}{h_{n-1}}\Phi_{n-1}(z),
\end{equation}
\begin{equation}
\frac{\partial \Phi_{n}^{*}(z^{-1})}{\partial t_{1}}
=\frac{h_{n}}{h_{n-1}}\Phi_{n-1}^{*}(z^{-1}),
\end{equation}
\begin{equation}
\frac{\partial \Phi_{n}^{*}(z^{-1})}{\partial \bar{t}_{1}}
=-\frac{S_{n}^{*}}{S_{n-1}^{*}}\frac{h_{n}}{h_{n-1}}
(\Phi_{n}^{*}(z^{-1})-z^{-1}\Phi_{n-1}^{*}).
\end{equation}
\eph
The compatibility condition gives the following nonlinear evolution equations:
\bph
\begin{equation}
\frac{\partial S_{n}}{\partial t_{1}}=S_{n+1}\frac{h_{n+1}}{h_{n}},
\;\;\;
\frac{\partial S_{n}}{\partial \bar{t}_{1}}=-S_{n-1}\frac{h_{n+1}}{h_{n}},
\label{11}
\end{equation}
\begin{equation}
\frac{\partial S^{*}_{n}}{\partial t_{1}}=-S_{n-1}^{*}\frac{h_{n+1}}{h_{n}},
\;\;\;
\frac{\partial S^{*}_{n}}{\partial \bar{t}_{1}}=S_{n+1}^{*}\frac{h_{n+1}}{h_{n}},
\label{22}
\end{equation}
\begin{equation}
\frac{\partial h_{n}}{\partial t_{1}}=S_{n}S_{n-1}^{*}h_{n},
\;\;\;
\frac{\partial h_{n}}{\partial \bar{t}_{1}}=S_{n}^{*}S_{n-1}h_{n}.
\label{rtc}
\end{equation}
\eph
(\ref{rtc})  is  exactly RTC written in somewhat  different form.\cite{m}
We can obtain two RTC's for the two times $t_{1}$ and $\bar{t}_{1}$.

Here we define  $a_{n}$, $b_{n}$ and $b_{n}^{*}$:
\bph
\begin{equation}
a_{n}\equiv 1-S_{n}S_{n}^{*}=\frac{h_{n+1}}{h_{n}},
\end{equation}
\begin{equation}
b_{n}\equiv S_{n}S_{n-1}^{*},
\end{equation}
\begin{equation}
b_{n}^{*}\equiv S_{n}^{*}S_{n-1}.
\end{equation}
\eph 

Notice that from  the definitions  $a_{n}$, $b_{n}$ and $b_{n}^{*}$
satisfy the periodic condition:
\begin{equation}
a_{n}=a_{n+N},\;\;\;
b_{n}=b_{n+N},\;\;\;
b_{n}^{*}=b_{n+N}^{*}.
\end{equation}

In terms of  $a_{n}$, $b_{n}$ and $b_{n}^{*}$,
 (\ref{11}) and (\ref{22}) become the two-dimensional  Toda  equations:
\bph
\begin{equation}
\frac{\partial a_{n}}{\partial t_{1}}
=a_{n}(b_{n+1}-b_{n}),\;\;\;
\frac{\partial b_{n}}{\partial \bar{t}_{1}}
=a_{n}-a_{n-1},
\label{tl1}
\end{equation}
and 
\begin{equation}
\frac{\partial a_{n}}{\partial \bar{t}_{1}}
=a_{n}(b_{n+1}^{*}-b_{n}^{*}),\;\;\;
\frac{\partial b_{n}^{*}}{\partial t_{1}}
=a_{n}-a_{n-1}.
\label{tl2}
\end{equation} 
\eph

This system has conjugate structures:
\begin{equation}
a_{n}\longrightarrow a_{n},
\;\;\;
b_{n}\longrightarrow  b_{n}^{*},
\;\;\;
b_{n}^{*}\longrightarrow  b_{n},
\;\;\;
t\longrightarrow \bar{t},
\;\;\;
\bar{t}\longrightarrow t.
\end{equation}
The two-dimensional Toda system with the conjugate structure 
can be seen is several physical models:
 the full unitary matrix models \cite{h} and $t\bar{t}$ fusion 
of the topological sigma models.\cite{cv},\cite{d}

We will assume that the eigenfunctions  $\Phi_{n}$ and $\Phi_{n}^{*}$ 
will obey the following  quasi-periodic conditions:
\begin{equation}
\Phi_{n+N}=\sqrt{z^{N}/\prod_{i=1}^{N}c_{i}}\Phi_{n},\;\;\;
\Phi_{n+N}^{*}=\sqrt{z^{-N}/\prod_{i=1}^{N}c_{i}^{*}}\Phi_{n}^{*},
\end{equation}
where
\begin{equation}
c_{n}=-\frac{S_{n}}{S_{n-1}}(1-S_{n}S_{n}^{*}),\;\;\;
c_{n}^{*}=-\frac{S_{n}^{*}}{S_{n-1}^{*}}(1-S_{n}S_{n}^{*}).
\end{equation}
If we set 
\begin{eqnarray}
& &
c_{n}=-f_{n}^{2},\;\;\;
c_{n}^{*}=-(f_{n}^{*})^{2},\;\;\;
z=k^{2},
\nonumber \\
& &
\Phi_{n}=k^{n}\alpha_{n}\varphi_{n},\;\;\;
\Phi_{n}^{*}=k^{n}\alpha_{n}^{*}\varphi_{n}^{*},\;\;\;
\alpha_{n-1}=\alpha_{n}f_{n},\;\;\;
\alpha_{n-1}^{*}=\alpha_{n}^{*}f_{n}^{*},
\end{eqnarray}
the new eigenfunctions  $\varphi_{n}$ and $\varphi_{n}^{*}$ 
will fulfill  periodic boundary condition, 
\begin{equation}
\varphi_{n+N}=\varphi_{n},\;\;\;
\varphi_{n+N}^{*}=\varphi_{n}^{*}.
\end{equation}
The linear problems are transformed into 
\begin{eqnarray}
& &(d_{n}-k^{2})\varphi_{n}+k(f_{n}\varphi_{n+1}+f_{n}\varphi_{n-1})
=0,
\nonumber \\
& &(d_{n}^{*}-k^{-2})\varphi_{n}^{*}+k(f_{n}^{*}\varphi_{n+1}^{*}
+f_{n}^{*}\varphi_{n-1}^{*})
=0,
\label{lf}
\end{eqnarray}
where 
\begin{equation}
d_{n}=-\frac{S_{n}}{S_{n-1}},\;\;\;
d_{n}^{*}=-\frac{S_{n}^{*}}{S_{n-1}^{*}}.
\end{equation}
We have the following  $N\times N$ matrix representation
of (\ref{lf})
\begin{equation}
L(k,\lambda)\varphi=0,\;\;\;
L^{*}(k,\lambda^{*})\varphi^{*}=0,
\end{equation}
where
\bph
\begin{eqnarray}
L=
\left(
\begin{array}{ccccc}
d_{1}-k^{2}& kf_{1}& & & kf_{N}\lambda\\
 kf_{1}&d_{2}-k^{2}&f_{2}& \bzr& \\
 & \cdots&\cdots& & \\
 & &\cdots&\cdots& \\
 & \bzr& & &kf_{N-1}\\
kf_{N}/\lambda& & & kf_{N-1}&d_{N}-k^{2}
\label{ma}
\end{array}
 \right),
\end{eqnarray}
\begin{eqnarray}
L^{*}=
\left(
\begin{array}{ccccc}
d_{1}^{*}-k^{-2}& k^{-1}f_{1}^{*}& & & k^{-1}f_{N}^{*}\lambda^{*}\\
 k^{-1}f_{1}^{*}&d_{2}^{*}-k^{-2}&f_{2}^{*}& \bzr& \\
 & \cdots&\cdots& & \\
 & &\cdots&\cdots& \\
 & \bzr& & &k^{-1}f_{N-1}^{*}\\
k^{-1}f_{N}^{*}/\lambda^{*}& & & k^{-1}f_{N-1}^{*}&d_{N}^{*}-k^{-2}
\label{ma*}
\end{array}
 \right),
\end{eqnarray}
\eph

Here $\lambda$ and $\lambda^{*}$ are the spectral parameters.
Notice that det$L(k,\lambda)$ and  det$L^{*}(k,\lambda^{*})$ are
 an integral  of motion 
for the two flows $t_{1}$ and $\bar{t}_{1}$.
The polynomial  det$L(k,\lambda)$ and  det$L^{*}(k,\lambda^{*})$
, thought being of degree
 $2N$ in $k$, have  only $N$ functionally independent coefficients 
respectively.
For $N$ even, it is an even function of $k$, while 
for odd, the only odd surviving
power of $k$ is the $N$th power.
Using  the (\ref{ic})  we can obtain the two curves ${\cal C}$ and 
${\cal C}^{*}$:
\bph
\begin{equation}
\epsilon^{2N}k^{N}(\lambda+\frac{1}{\lambda})
=k^{2N}+I_{N-1}k^{2(N-1)}+\cdots+
I_{1}k^{2}\pm1,
\end{equation}
\begin{equation}
\epsilon^{2N}k^{-N}(\lambda^{*}+\frac{1}{\lambda^{*}})
=k^{-2N}+I_{N-1}^{*}k^{-2(N-1)}+\cdots+
I_{1}^{*}k^{-2}\pm1.
\end{equation}
\eph
For $N$ even, the sign of the last term is $+$,
 while for odd, it is $-$.
If we set $k=e^{x}$, 
then we can obtain
\bph
\begin{equation}
2^{-N}\epsilon^{2N}(\lambda+\frac{1}{\lambda})
=
\prod_{i}[\sinh(x-\hat{\alpha}_{i})],
\label{c1}
\end{equation}
\begin{equation}
2^{-N}\epsilon^{2N}(\lambda^{*}+\frac{1}{\lambda^{*}})
=
\prod_{i}[\sinh(-x-\hat{\alpha}_{i}^{*})],
\label{c2}
\end{equation}
\eph
where 
\begin{equation}
\sum_{i}\hat{\alpha}_{i}=0, \;\;\;\sum_{i}\hat{\alpha}_{i}^{*}=0.
\end{equation}

The two curves have ${\rm Z}_{N}$ and ${\rm Z}_{2}$  symmetry respectively,\cite{n}
\begin{equation}
\hat{\alpha}_{i}\rightarrow \hat{\alpha}_{i}+\frac{2\pi {\rm i}k}{N},\;\;\;
\hat{\alpha}_{i}^{*}\rightarrow \hat{\alpha}_{i}^{*}+\frac{2\pi {\rm i}l}{N},\;\;\;\;\;
I_{n}\rightarrow e^{-2\pi {\rm i}k(N-n)/N}I_{n},\;\;\;
I_{n}^{*}\rightarrow e^{-2\pi {\rm i}l(N-n)/N}I_{n}^{*},
\label{zn}
\end{equation}
and 
\begin{equation}
\lambda\rightarrow -\lambda,\;\;\;
\lambda^{*}\rightarrow -\lambda^{*},\;\;\;
f_{n}\rightarrow -f_{n},\;\;\;
f_{n}^{*}\rightarrow -f_{n}^{*}.
\end{equation}
\setzero
\section{The Whitham-Toda equations}

In this section we consider the two-dimensional 
Toda lattice system (\ref{tl1}) and (\ref{tl2})
without the periodic condition.
This system can be embedded in the 
two-dimensional Toda hierarchy.\cite{m}
The matrix version of the spectral problem is
\begin{equation}
{\cal L}_{nk}\phi_{k}(z)=z\phi_{n}(z),\;\;\;
{\cal L}_{nk}^{*}\phi_{k}^{*}(z)=z^{-1}\phi_{n}^{*}(z),
\end{equation}
\begin{eqnarray}
\frac{\partial \phi_{n}(z)}{\partial t_{m}}
&=&
-[({\cal L}^{m})_{-}]_{nk}\phi_{k}(z),\;\;\;
\frac{\partial \phi_{n}(z)}{\partial \bar{t}_{m}}
=
[({\cal L}^{*m})_{-}]_{nk}\phi_{k}(z),
\nonumber \\
\frac{\partial \phi_{n}^{*}(z)}{\partial t_{m}}
&=&
[({\cal L}^{m})_{+}]_{nk}\phi_{k}^{*}(z),\;\;\;
\frac{\partial \phi_{n}^{*}(z)}{\partial \bar{t}_{m}}
=
-[({\cal L}^{*m})_{+}]_{nk}\phi_{k}^{*}(z),\;\;\;
m=1,2,\cdots
\end{eqnarray}
where 
$(A)_{+}$ is the upper triangular part of of the matrix $A$
 (including the main diagonal)
 while  $(A)_{-}$ is strictly the lower triangular part.
$\cal{L}$ and $\cal{L}^{*}$ are the Lax operators 
and $\phi_{n}(z)$ and $\phi_{n}^{*}(z)$  are the Baker-Akhiezer (BA)
 functions.

The two-dimensional Toda lattice system
(\ref{tl1}) and (\ref{tl2})
has infinite number of conserved quantities.
The   conserved density is  obtained 
from the Lax operators
\begin{eqnarray}
D_{1}^{T}&=&b_{n},\;\;\;\bar{D}_{1}^{T}=b_{n}^{*},
\nonumber \\
D_{2}^{T}&=&\frac{1}{2}b_{n}^{2}-\frac{a_{n}}{1-a_{n}}b_{n}b_{n-1}
,\;\;\;
\bar{D}_{2}^{T}=\frac{1}{2}b_{n}^{*2}-\frac{a_{n}}{1-a_{n}}
b_{n}^{*}b_{n-1}^{*}.
\label{tc}
\end{eqnarray}

The conservation laws are given by
\bph
\begin{equation}
\frac{\partial D_{i}}{\partial t_{k}}
=
(\Delta-1)F_{i},\;\;\;
\frac{\partial D_{i}}{\partial \bar{t}_{k}}
=
(\Delta-1)\tilde{F}_{i},
\end{equation}
\begin{equation}
\frac{\partial D^{*}_{i}}{\partial \bar{t}_{k}}
=
(\Delta-1)F^{*}_{i},\;\;\;
\frac{\partial D_{i}^{*}}{\partial t_{k}}
=
(\Delta-1)\tilde{F}^{*}_{i},
\end{equation}
\eph
where 
$\Delta$ is the unit shift operator,
 i.e., $\Delta f_{n}=f_{n+1}$,
 and $F_{i}$ and $\tilde{F}_{i}$ are  the flows
for the time $t_{1}$ and $\bar{t}_{1}$. 
For example we can obtain 
\begin{eqnarray}
F_{1}&=&F_{1}^{*}=a_{n},
\nonumber
\\
\tilde{F}_{1}&=&\frac{a_{n}}{1-a_{n}}b_{n}b_{n-1},\;\;\;
\tilde{F}_{1}^{*}=\frac{a_{n}}{1-a_{n}}b_{n}^{*}b_{n-1}^{*},
\nonumber \\
F_{2}&=&a_{n}b_{n},\;\;\;F_{2}^{*}=a_{n}b_{n}^{*},
\nonumber \\
\tilde{F}_{2}&=&
\frac{a_{n}}{1-a_{n}}\frac{a_{n-1}}{1-a_{n-1}}b_{n}b_{n-1}b_{n-2}
x-\frac{a_{n}}{1-a_{n}}b_{n}^{2}b_{n-1},
\nonumber 
\\
\tilde{F}_{2}^{*}&=&
\frac{a_{n}}{1-a_{n}}\frac{a_{n-1}}{1-a_{n-1}}b_{n}^{*}b_{n-1}^{*}b_{n-2}^{*}
-\frac{a_{n}}{1-a_{n}}b_{n}^{*2}b_{n-1}^{*}.
\label{fl}
\end{eqnarray}
The $t_{1}$-flow and $\bar{t}_{1}$-flow  of the conserved density  are different.
There is the inhomogeneity  between 
$t_{k}$ and $\bar{t}_{k}$.

Here we introduce the  1-form $dS$ and $dS^{*}$
 which are meromorphic on ${\cal C}$ and ${\cal C}^{*}$.
\begin{equation}
d S\cong\log z \frac{d\lambda}{\lambda},
\;\;\;\;\;
d S^{*}\cong\log z \frac{d\lambda^{*}}{\lambda^{*}}.
\end{equation}

To obtain the Whitham equations, we introduce two time scales 
and then average the flux and density  over the  fast variables
 in the conservation laws.
We break up the dynamics into slow and  fast scales.
Let $n$, $t_{k}$ and $\bar{t}_{k}$ denote the fast   spatial  and time 
variables and let $X=\epsilon x$, $T_{k}=\epsilon t_{k}$ 
and $\bar{T}_{k}=\epsilon t_{k}$.

\bph
\begin{equation}
\frac{\partial }{\partial T_{k}}dS
=
d\Omega_{k},\;\;\;
\frac{\partial }{\partial \bar{T}_{k}}dS
=
d\tilde{\Omega}_{k},\;\;\;
\frac{\partial }{\partial X}dS
=
d\Omega_{0},
\label{dS}
\end{equation}
and 
\begin{equation}
\frac{\partial }{\partial \bar{T}_{k}}dS^{*}
=
d\Omega_{k},\;\;\;
\frac{\partial }{\partial T_{k}}dS^{*}
=
d\tilde{\Omega}_{k},\;\;\;
\frac{\partial }{\partial X}dS^{*}
=
d\Omega_{0}^{*}.
\label{dS*}
\end{equation}
\eph

$d\Omega_{0}$, $d\Omega_{k}$, $d\Omega_{0}^{*}$ and $d\Omega_{k}^{*}$
 are the normalized  Abelian  differentials.
These  are normalized by the conditions 
\begin{equation}
\oint_{\alpha_{i}}d\Omega_{0}= \oint_{\alpha_{i}}d\Omega_{k}
=\oint_{\alpha_{i}^{*}}d\Omega^{*}_{0}
=\oint_{\alpha_{i}^{*}}d\Omega_{k}^{*}=0,
\end{equation}
where
$\alpha_{i}$and $\alpha_{i}^{*}$ are the standard symplectic basis 
of homology cycles of the curve (\ref{c1}) and (\ref{c2}).

We can write  the Whitham equations in the terms of 
the Aberian  differentials for the general $T_{k}$ and $\bar{T}_{k}$ 
as  follows 
\bph
\begin{equation}
\frac{\partial }{\partial T_{k}}d\Omega_{0}
=
\frac{\partial }{\partial X}d\Omega_{k},\;\;\;
\frac{\partial }{\partial \bar{T}_{k}}d\Omega_{0}
=
\frac{\partial }{\partial X}d\tilde{\Omega}_{k}.
\label{cl1}
\end{equation}
and 
\begin{equation}
\frac{\partial }{\partial \bar{T}_{k}}d\Omega^{*}_{0}
=
\frac{\partial }{\partial X}d\Omega_{k}^{*},\;\;\;
\frac{\partial }{\partial T_{k}}d\Omega_{0}^{*}
=
\frac{\partial }{\partial X}d\tilde{\Omega}_{k}^{*}.
\label{cl2}
\end{equation}
\eph
This  equation is   an average  form of the  conservation law.

\setzero
\section{Five-Dimensional Gauge Theory}

The prepotential ${\cal F}$  is identified  with logarithm of the
 $\tau$-function of the Whitham hierarchy.
The system which we considered in the section 2 and 3
has one $\tau$-function but
has the two curves.
To describe the five-dimensional moduli space
 the two curves have to be unified.

The Coulomb branch of the moduli space  is given by 
$\Psi={\rm diag}(\hat{\alpha}_{1},\hat{\alpha}_{2},\cdots,\hat{\alpha}_{N})$ 
with
$\sum_{i}\hat{\alpha}_{i}=0$, modulo the Weyl  group action, which permutes the 
$\hat{\alpha}_{i}$.
It can thus be taken to be the Weyl chamber 
$\hat{\alpha}_{1}\geq \hat{\alpha}_{2}\geq\cdots\geq \hat{\alpha}_{N}$. 
To unify $ {cal C}$ (\ref{c1}) and ${cal C}^{*}$ (\ref{c2})
 we  set  as follows:
\begin{equation}
\lambda^{*}=\pm \lambda,\;\;\;
\hat{\alpha}_{i}^{*}=-\hat{\alpha}_{N-i+1},\;\;\;
I_{n}^{*}=I_{N-n+1}.
\label{sy}
\end{equation}  
For $N$ even, the sign  is $+$,
 while for $N$ odd, it is $-$.

The effective prepotential 
on the Coulomb  branch
 has appeared in 
\cite{i} and \cite{n}.
The prepotential 
is invariant under   the $Z_{2}$ transformation:
$\hat{\alpha}\rightarrow -\hat{\alpha}$.
The gauge group $SU(N)$ 
is broken by the Higgs mechanism.
However the Weyl group should still be unbroken, 
so  this $Z_{2}$ symmetry resides.
In the physical meaning 
(\ref{sy})  is the charge conjugation.\cite{i}

From this conjugation we can obtain next relations:
\begin{equation}
dS=dS^{*},\;\;\;
d\Omega_{0}=d\Omega_{0}^{*},\;\;\;
d\Omega_{k}=d\tilde{\Omega}_{k}^{*}
\end{equation}
This result means the homogeneity 
between $T_{k}$ and $\bar{T}_{k}$.
But  the conserved quantity is exchanged as (\ref{sy}).  
For the $SU(2)$ case the conjugate motions are the same, 
as $I_{2}=I_{2}^{*}$.
Then the set of the holomorphic differentials 
 become as follows:
\begin{equation}
d\omega_{n}
 =d\omega_{N-n-1}^{*},
\end{equation}
where
\begin{equation}
d\omega_{k}=\frac{\partial dS}{\partial I_{k}},
\;\;\;\;\;\;\;\;\;\;
d\omega_{k}^{*}=\frac{\partial dS^{*}}{\partial I_{k}^{*}}.
\end{equation}
\setzero
\section{Concluding Remarks}

We study the five-dimensional supersymmetric SU(N) gauge theory 
and the relativistic Toda (RTC) equations.
RTC is described as a particular reduction of the two-dimensional 
Toda lattice hierarchy.
This system has two curves.
To describe the  moduli space of the five-dimensional gauge theory,
 the two curves have to be unified.
Then the two times $T_{k}$ and $\bar{T}_{k}$ are homogeneous 
and the conjugate structure becomes the charge conjugation.

\end{document}